\documentclass{PoS}

\title{Effects of dispersive wave modes on charged particles transport}

\ShortTitle{Effects of dispersive wave modes on charged particles transport}

\author{\speaker{Cedric Schreiner}$^{ab}$ and Felix Spanier$^a$\\
        $^a$Center for Space Research, North-West University Potchefstroom, South Africa\\
	$^b$Lehrstuhl f\"ur Astronomie, Julius-Maximilians-Universit\"at W\"urzburg, Germany\\
        E-mail: \email{cschreiner@astro.uni-wuerzburg.de}, \email{felix@fspanier.de}}

\abstract{The transport of charged particles in the heliosphere and the interstellar medium is governed by the interaction of particles and magnetic irregularities.
For the transport of protons a rather simple model using a linear Alfv\'{e}n wave spectrum which follows the Kolmogorov distribution usually yields good results.
Even magnetostatic spectra may be used.
For the case of electron transport, particles will resonate with the high-k end of the spectrum.
Here the magnetic fluctuations do not follow the linear dispersion relation, but the kinetic regime kicks in.\\
We will discuss the interaction of fluctuations of dispersive waves in the kinetic regime using a particle-in-cell code \cite{kilian12}.
Especially the scattering of particles following the idea of Lange et al. \cite{lange13} and its application to PiC codes \cite{schreiner14} will be discussed.
The effect of the dispersive regime on the electron transport will be discussed in detail.}

\FullConference{The 34th International Cosmic Ray Conference,\\
		30 July- 6 August, 2015\\
		The Hague, The Netherlands}

\begin{document}

\section{Introduction}
\label{sec:intro}

The solar wind plasma can be understood as a mainly thermal population of charged particles, substantially protons and electrons, with an excess of highly energetic particles.
To understand many of the phenomena found in the solar wind it is crucial to study these energetic particles and the processes involved in their acceleration and transport.
The transport characteristics of charged particles in the solar wind plasma is dominated by the interaction with magnetic fields and inhomogeneities.
However, the complete system of electromagnetic fields and charged particles is highly nonlinear and thus it is difficult to find a theoretical approach that is able to model these complex processes.
Analytical approximations describing the interaction of protons and plasma waves in the magneto-hydrodynamic (MHD) regime have been found for a linearized version of the problem \cite{jokipii66,lee74,schlickeiser89}, but with the ever increasing amount of computing power available numerical simulations have also become a standard tool for studying particle scattering, although different methods can be used \cite{lange13,michalek96,qin01}.
Nonetheless most numerical simulations cover only the interaction of protons with magnetic fluctuations.
With our work we attempt to transfer the approach to analyzing proton scattering with fully kinetic particle-in-cell (PiC) simulations \cite{schreiner14} to a setup where the scattering of electrons and dispersive waves can be studied.

\section{Theory}
\label{sec:theory}

A full description of a magnetized plasma can be achieved using the relativistic Vlasov equation, which yields a statistical description of the particle transport.
It is convenient to assume a collisionless plasma for the description of the solar wind, so a particle's motion is affected only by the Lorentz force.
In this framework the quasi-linear theory (QLT), which was first introduced by Jokipii \cite{jokipii66}, can be used to study scattering processes of particles and plasma waves.
QLT describes these processes as resonant interactions of particles and the electromagnetic fields of the wave, where resonant scattering leads to a drastic change in the particle's direction of motion.
The resonance condition in the rest frame of the plasma \cite{schlickeiser89}
\begin{equation}
	\omega - k_\parallel \, \mu \, v_\alpha = n \, \Omega_\alpha / \gamma_\alpha
	\label{resonance}
\end{equation}
for a resonance of order $n$ yields $\mu$ as the cosine of the resonant pitch angle $\theta$ -- i.e. the angle between the particle's direction of motion and the background magnetic field $\vec{B}_0$ in the plasma -- for which the scattering process is most efficient.
In the above equation $\omega$ is the wave's frequency and $k_\parallel$ the component of its wave vector which is parallel to $\vec{B}_0$.
For a particle of species $\alpha$ the speed, Lorentz factor and gyrofrequency are denoted by $v_\alpha$, $\gamma_\alpha$ and $\Omega_\alpha = (q_\alpha \, B_0) / (m_\alpha \, c)$, respectively, with the particle's charge $q_\alpha$, its mass $m_\alpha$ and the speed of light $c$.
\newline
Additional to the position of the resonance in terms of the resonant pitch angle $\mu_{res}$, the scattering amplitude $\Delta\mu(\mu)$ can also be studied using QLT predictions.
An analytical approximation of the scattering amplitude at a given time $t$ after the onset of the interaction is provided by \cite{lange13} (for order $|n|=1$):
\begin{equation}
	\Delta\mu^{'\pm}(\mu',t,\Psi^\pm) = \frac{\Omega_\alpha}{\gamma'_\alpha} \frac{\delta B}{B_0} \sqrt{1-\mu^{'2}} \cdot \frac{\cos{\left(\Psi^\pm\right)} - \cos{\left((\pm k_\parallel \, v'_\alpha \, \mu' - \Omega_\alpha / \gamma'_\alpha) \cdot t + \Psi^\pm\right)}}{\pm k_\parallel \, v'_\alpha \, \mu' - \Omega_\alpha / \gamma'_\alpha}.
	\label{qlt-deltamu}
\end{equation}
Here, $\Psi^\pm$ is the phase angle which maximizes $\Delta\mu'$ and which is defined by
\begin{equation}
	\Psi^\pm(\mu',t) = \arctan{\left(\frac{\sin{\left((\pm k_\parallel \, v'_\alpha \, \mu' - \Omega_\alpha / \gamma'_\alpha) \cdot t\right)}}{1 - \cos{\left((\pm k_\parallel \, v'_\alpha \, \mu' - \Omega_\alpha / \gamma'_\alpha) \cdot t\right)}}\right)}.
	\label{qlt-psi}
\end{equation}
Equations (\ref{qlt-deltamu}) and (\ref{qlt-psi}) are formulated in the rest frame of the wave, denoted by the primed parameters, and only hold if the magnetic field of the wave $\delta B$ is small compared to the background magnetic field $B_0$.
The $(+)$ solution describes the scattering of particles and waves with right handed polarization, whereas the $(-)$ solution describes the interaction with left handed waves.
To obtain the full set of solutions $\Delta\mu^{'\pm}(\mu',t,\Psi^\pm)$ and $\Delta\mu^{'\pm}(\mu',t,\Psi^\pm+\pi)$ have to be evaluated.
\newline
Since equations (\ref{qlt-deltamu}) and (\ref{qlt-psi}) are formulated in the co-moving frame of the wave and particle velocities and pitch angles are measured in the rest frame of the plasma the primed parameters have to be expressed by their non-primed counterparts.
The simple transformation
\begin{equation}
	\mu' = \mu - \frac{\omega}{k\,v_\alpha},
	\label{mutransform}
\end{equation}
used by \cite{lange13,schreiner14} is valid if the phase speed of the wave $v_{ph} = \omega/k$ is small compared to the speed of the particle -- assuming parallel propagating waves with $k_\parallel = k$.
This is normally true for low frequency waves such as the dispersionless Alfv\'en waves.
But since we are discussing dispersive waves, which might have higher phase speeds, we will use a different and more exact transformation.
In the plasma frame, the pitch angle is defined by
\begin{equation}
	\mu = \frac{\vec{v}_\alpha \cdot \vec{B}_0}{v_\alpha \, B_0},
	\label{pitch-plasma}
\end{equation}
where $v_\alpha$ and $B_0$ are the absolute values of $\vec{v}_\alpha$ and $\vec{B}_0$.
Analogously, the pitch angle can be defined in the wave frame by
\begin{equation}
	\mu' = \frac{\vec{v}'_\alpha \cdot \vec{B}_0}{v'_\alpha \, B_0} = \frac{\left(\vec{v}_\alpha - \vec{v}_{ph}\right) \cdot \vec{B}_0}{\left|\vec{v}_\alpha - \vec{v}_{ph}\right| \, B_0}.
	\label{pitch-wave}
\end{equation}
Assuming a parallel propagating wave ($\vec{v}_{ph} \parallel \vec{B}_0$) the velocity $\vec{v}'_\alpha$ can be expressed as
\begin{eqnarray}
	\vec{v}'_\alpha &=& \left(v_\alpha \, ~ \mu - v_{ph}, v_\alpha \, \sqrt{1-\mu^2} \, \cos{(\Phi)}, ~ v_\alpha \, \sqrt{1-\mu^2} \, \sin{(\Phi)}\right),
	\label{vel_wave}
	\\
	v'_\alpha &=& \sqrt{v_\alpha^2 - 2 \, v_\alpha \, v_{ph} \, \mu + v_{ph}^2},
	\label{vel_wave_abs}
\end{eqnarray}
with the polar angle $\Phi$ of the particle\footnote{Note that this is only applicable in the case of weakly relativistic velocities.}.
Using these expressions equation (\ref{pitch-wave}) becomes
\begin{equation}
	\mu' = \frac{\mu v_\alpha - v_{ph}}{\sqrt{v_\alpha^2 - 2 \, v_\alpha \, v_{ph} \, \mu + v_{ph}^2}}.
	\label{mu_wave}
\end{equation}
Finally, we can write the Lorentz factor in the wave frame as
\begin{equation}
	\gamma'_\alpha = \left(\sqrt{1 - \frac{v_\alpha^2 - 2 \, v_\alpha \, v_{ph} \, \mu + v_{ph}^2}{c^2}}\right)^{-1}.
	\label{gamma_wave}
\end{equation}
Inserting (\ref{vel_wave_abs}), (\ref{mu_wave}) and (\ref{gamma_wave}) into equations (\ref{qlt-deltamu}) and (\ref{qlt-psi}) allows us to predict the time evolution of the scattering amplitudes in the plasma frame.
\newline
The equations presented above are only valid under the condition of parallel propagating, purely transverse waves with a weak magnetic field compared to the background magnetic field.
Thus we will limit our model to parallel propagating waves in the low frequency regime, namely Alfv\'en and Whistler waves.
The dispersion relations for those waves can be derived in the approximation of a cold plasma \cite{stix92} and are given by
\begin{equation}
	|k^\pm| = \frac{\omega}{c} \sqrt{1 - \frac{\omega_p^2}{(\omega \pm \Omega_e) (\omega \pm \Omega_p)}},
	\label{disp}
\end{equation}
in the case of a plasma consisting only of protons and electrons.
Here, $k^+$ denotes the wave mode with right handed and $k^-$ the mode with left handed polarization.
In equation (\ref{disp}) $\omega_p$ is the plasma frequency and $\Omega_p$ and $\Omega_e$ are the gyrofrequencies of protons and electrons, respectively.
Note that the electron gyrofrequency $\Omega_e$ is defined with a negative sign.
In their low frequency regimes ($\omega < \Omega_p$ or $\omega < |\Omega_e|$), $k^-$ and $k^+$ contain the Alfv\'en and Whistler modes, respectively.
\newline
Both wave modes start with a dispersionless regime at frequencies $\omega \ll \Omega_p$ where the dispersion relation can be approximated by $\omega = k \, v_A$, where $v_A$ is the Alfv\'en speed.
At higher frequencies the phase speed of the Alfv\'en waves decreases with increasing frequency until the wave mode reaches a cyclotron resonance at $\Omega_p$.
For Whistler waves the phase speed first increases with increasing frequency and then decreases again when the wave mode approaches cyclotron resonance at $|\Omega_e|$.

\section{Particle-in-cell approach}
\label{sec:pic}

\subsection{Code overview}
\label{sec:overview}

For our numerical simulations of processes in the solar wind plasma we employ a particle-in-cell (PiC) approach and use our code \emph{ACRONYM} (\emph{A}dvanced \emph{C}ode for \emph{R}elativistic \emph{O}bjects, \emph{N}ow with \emph{Y}ee-Lattice and \emph{M}acroparticles) \cite{kilian12}.
The basic idea of the PiC method is to solve the coupled system of Vlasov equation and Maxwell's equations by using phase space samples, so-called macro particles, to represent charged particles in a plasma \cite{hockney88}.
These (macro) particles induce electromagnetic fields which then act back on the charged particles as they move.
This is achieved by calculating the currents which are created by the particles' motion and feeding them into Maxwell's equations, where they serve as source terms.
Maxwell's equations are solved on a grid and thus currents, charge densities and the electromagnetic fields are stored only at discrete grid positions.
The fields can be interpolated between grid points to obtain the Lorentz force at the positions of the particles.
In this manner the PiC method allows self-consistent simulations of collisionless plasmas with arbitrary initial and boundary conditions.
\newline
\emph{ACRONYM} in particular is an explicit, electromagnetic and fully relativistic PiC code, which has been developed by members of our working group.
The code is capable of simulating kinetic, collisionless, magnetized plasmas self-consistently in a full 3d3v setup, which means that three spatial dimensions are resolved and that electromagnetic fields as well as particle velocities are treated as three dimensional vectors.
For testing purposes or for saving huge amounts of computing time 1d3v and 2d3v versions are also available, although their applicability depends on the physical problem to be simulated.
The concept of the code is to maintain second order accuracy in both space and time, which is achieved by the combination of a suitable set of algorithms\cite{kilian12}.

\subsection{Application to wave particle scattering}
\label{sec:application}

The scattering of particles and low-frequency waves can be analyzed using magneto-hydro-dynamic (MHD) simulations and test particles, which only react to the electromagnetic fields produced by the MHD simulation.
This approach is very convenient for the case of dispersionless Alfv\'en waves acting on protons at very low frequencies $\omega \ll \Omega_p$.
A direct comparison of MHD simulations by Lange et al. \cite{lange13} and equivalent PiC simulations performed using the \emph{ACRONYM} code was presented in \cite{schreiner14}.
The article demonstrates that even with the assumption of an artificial mass ratio $m_p / m_e$ and optimized, unrealistic parameters PiC simulations require far more computational resources than full-scale MHD simulations with solar wind parameters.
\newline
Still we stick to the PiC approach, since PiC codes are much more versatile than MHD codes.
For our purposes the key advantage of PiC over MHD is that in a PiC simulation all kinds of physically possible plasma waves can be generated, whereas incompressible MHD is limited to Alfv\'en waves only.
We have shown that resonant scattering processes of protons and left handed waves in the dispersive regime of the Alfv\'en mode are possible \cite{schreiner14} and we have also presented proof of concept simulations suggesting that high energetic protons are able to interact with right handed waves in the low-frequency end of the dispersive regime of the Whistler mode\footnote{C. Schreiner, U. Ganse, F. Spanier, \emph{Wave-Particle-Interaction in Kinetic Simulations}, oral presentation at \emph{CHPC National Meeting 2014}, Kruger National Park, South Africa, 01.12.2014 - 05.12.14, to appear online at http://www.chpcconf.co.za/index.php/presentations.}.
\newline
A more suitable problem for PiC simulations is the scattering of electrons and waves, since these interactions happen on smaller time and length scales.
PiC codes have to resolve the Debye length and time steps satisfying the CFL-condition -- scales which are comparable to electron scales.
Thus, less computational effort is needed since the simulation requires less grid cells and less time steps to cover the typical scales of electron driven processes than it would need to resolve proton scales.
\newline
To analyze electron scattering we follow the model setup of Lange et al. \cite{lange13} which we have adapted for \emph{ACRONYM} \cite{schreiner14}:
A single wave mode is excited at the beginning of the simulation by initializing the electric and magnetic fields in such a way that they represent a wave with a specific wave length and amplitude and that the polarization of the wave is reproduced correctly.
The properties of the wave are derived in cold plasma approximation following \cite{stix92}, which is sufficient for relatively low plasma temperatures (i.e. non-relativistic thermal velocities) although the simulation itself also treats thermal effects.
We then perform an initial particle boost, which superposes an ordered motion following the wave's fields to the random thermal motion of each particle.
This ensures that the wave is driven in the right direction and that it does not lose energy to the particles by having to accelerate them at the beginning of the simulation.
The initial excitation of a single wave yields a magnetized background plasma which is governed by the dynamics of this particular wave.
The whole spectrum of other waves, which is excited thermally during the first few time steps of the simulation, can be neglected when looking at resonant scattering of particles and waves.
\newline
Scattering properties are studied by following the trajectories of test particles which are initialized as a separate population of either electrons or protons.
The only difference between test and background particles is that test particles have a macro factor of one, which means that each of them represents only a single, physical particle.
For simplicity and to be able to analyze their scattering behavior more precisely test particles are initialized mono-energetically and with isotropic angular distribution in velocity space.
This makes sure that the whole range of pitch angles $-1 \leq \mu \leq 1$ is covered.
By choosing the appropriate speed for the particles the resonant pitch angle $\mu_{res}$ can be selected.
During the initialization of the simulation the resonance condition (\ref{resonance}) is solved for the velocity $v_\alpha$ in the plasma frame:
\begin{equation}
	v_\alpha = \left|\frac{k_\parallel \, \omega \, |\mu_{res}| \pm |\Omega_\alpha| \, \sqrt{k_\parallel^2 \, \mu_{res}^{2} + \frac{\Omega_\alpha^2 - \omega^2}{c^2}}}{k_\parallel^2\,\mu_{res}^{2} + \frac{\Omega_\alpha^2}{c^2}}\right|.
	\label{vel}
\end{equation}
Note that the order of the resonance has been set to $|n|=1$ in the above equation and that $\mu_{res}$ denotes the resonant pitch angle in the plasma frame.
Only one of the two solutions of equation (\ref{vel}) satisfies the resonance condition (\ref{resonance}).
This has to be checked in a separate step.
\newline
During the simulation the pitch angles of all test particles are calculated and written to disk during specific output time steps.
To study the scattering processes we then simply analyze the changes in the particles' pitch angles:
\begin{equation}
	\Delta\mu = \mu(t) - \mu(t_0),
	\label{deltamu-sim}
\end{equation}
where $t$ is the current time in the simulation at a specific output time step and $t_0 = 0$ marks the beginning of the simulation.

\section{Simulations}
\label{sec:sims}

\begin{figure}[h]
	\centering
	\includegraphics[width=1.0\linewidth]{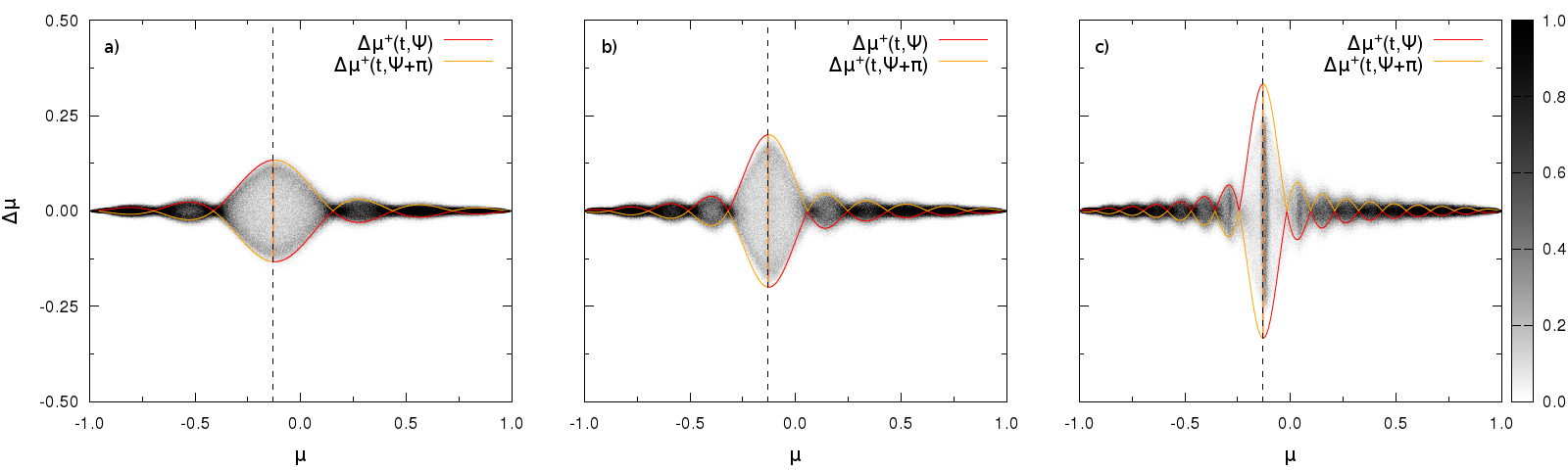}
	\caption{
		Resonant scattering of 100 keV electrons and a single Whistler wave.
		Panels a), b) and c) show the scattering behavior of electrons after \mbox{$4.0\cdot10^5$}, \mbox{$6.0\cdot10^5$} and \mbox{$1.0\cdot10^6$} time steps, respectively.
		Simulation data (gray scale) and QLT predictions (red and yellow curves, dashed line) agree well throughout the simulation.
		Color coding represents test particle number density $n_e(\mu,\Delta\mu)$, normalized to 1/10000 of the total number of test electrons.
	}
	\label{fig:scatter}
\end{figure}

As demonstrated in our previous work \cite{schreiner14} simulations with realistic parameters representing the actual circumstances in the solar wind are not feasible with PiC when proton scattering is considered.
However, the simulation of electron scattering requires much less numerical effort.
Thus we try to employ parameters which are at least close to those in the solar wind.
To estimate typical quantities which are relevant for the plasma, such as background magnetic field $B_0$, temperature $T$ or plasma frequency $\omega_p$ at a specific distance to the sun, we follow the analytical model of Vainio et al. \cite{vainio03}.
For first tests we tweak those parameters in order to decrease the computational effort for the simulation or to achieve a setup which is easily comparable to theoretical predictions.
\newline
\begin{table}[h]
	\centering
	\begin{tabular}{|c|c|c|c|c|c|c|c|}
		\hline
		& $B_0$ (G) & $\omega_p$ (rad/s) & $v_{th,e}$ (c) & $k$ (cm$^{-1}$) & $\omega$ (rad/s) & $\delta B$ (G)
		\\
		\hline
		model & $0.20$ & $6.4\cdot10^7$ & $2.0\cdot10^6$ & -- & -- & --
		\\
		\hline
		simulation & $0.20$ & $1.5\cdot10^8$ & $2.0\cdot10^4$ & $5.9\cdot10^{-3}$ & $2.1\cdot10^6$ & $2.0\cdot10^{-3}$
		\\
		\hline
	\end{tabular}
	\caption{Parameter set for test particle simulation (see text for details). Model parameters from \cite{vainio03} are computed for a distance of $2.6$ solar radii from the sun.}
	\label{tab:paras}
\end{table}
\newline
In this article we would like to present a first 2d3v simulation in which the interaction of fast electrons with a kinetic energy of \mbox{$E_{kin} \approx 100$ keV} with a right handed, circularly polarized wave in the dispersive regime of the Whistler branch is modeled.
The key parameters of the simulation are listed in table \ref{tab:paras}.
We start with a background magnetic field of \mbox{$B_0 = 0.20$ G} which corresponds to the field strength at $2.6$ solar radii from the sun according to \cite{vainio03}.
However we increase the plasma frequency by a factor of about $2.3$ -- and thus the density by a factor of $5.5$ -- and decrease the temperature by a factor of 100, compared to the model parameters at $2.6$ solar radii.
As stated above, this is done not to reflect a specific region in the solar wind, but to optimize the parameters for first numerical tests.
\newline
The simulation is arranged in a way that minimizes computational effort.
The numerical box size of \mbox{$4096 \times 128$ grid} cells resolves the gyration of thermal electrons (short edge) and one wave length of the amplified wave (long edge).
We have chosen the minimal wave length possible, meaning that Whistler waves at smaller wave lengths -- or larger wave numbers $k$ -- are dissipated by cyclotron damping.
With a duration of $1\cdot10^6$ time steps the simulation covers the initial phase in which the wave particle resonance is developed and the evolution of the interaction into a state of constant scattering amplitude.
Over this period of time the results from the simulation can be compared to QLT prediction.
The simulation is initialized with an equal number of background protons and electrons, ten of each species per cell.
Additionally, two test electrons per cell -- with a macro factor of one -- are included, which adds up to about one million test particles in the whole simulation.
\newline
We store the pitch angles of each test electron at specific time steps and then can plot the change of the particle's pitch angle $\Delta\mu$ as defined in equation (\ref{deltamu-sim}) over its time averaged pitch angle \mbox{$\bar{\mu} = (\mu(t) + \mu(t_0))/2$}.
Figure \ref{fig:scatter} shows such scatter plots at different time steps in the simulation.
Additionally to the simulation data we show QLT predictions from equations (\ref{qlt-deltamu}) and (\ref{qlt-psi}) -- with corresponding transformations into the plasma frame as described in section \ref{sec:theory}.
\newline	
The three panels in figure \ref{fig:scatter} show the evolution of the scattering amplitude $\Delta\mu(\mu)$ over time.
In accordance with QLT predictions the resonance grows in amplitude while its width decreases (panels a) and b)).
The position of the resonance is given by $\mu_{res}^+ = (\omega + \Omega_e / \gamma_e) / (k \, v_e)$ and marked by the dashed lines in the scatter plots.
The physical time simulated during the $1\cdot10^6$ numerical time steps represents roughly $2.8 ~ T_e$, where $T_e = 2 \, \pi/\Omega_e$ is the gyration timescale for electrons.
At the end of the simulation the resonance is already fully developed.
The amplitude of the resonance peak stays constant at $\Delta\mu_{max} \sim 0.25$, while QLT predicts a further increase (panel c)).
However, the predicted increase is not physical and marks the transition region where QLT becomes invalid, since equation (\ref{qlt-deltamu}) develops a $\delta$-shaped resonance with infinite amplitude for $t\rightarrow\infty$.

\section{Outlook}
\label{sec:outlook}

Our first test simulation shows that the PiC approach allows to simulate resonant scattering of fast electrons and Whistler waves self-consistently and in accordance with theoretic predictions.
Although the set of physical parameters chosen for this simulation does not represent actual conditions in the solar wind we are confident that such simulations are able to contribute to our understanding of kinetic processes in the heliosphere.
As a next step a physical regime in the solar wind has to be found which is accessible to PiC simulations with reasonable computational effort.
But even if a realistic setup is out of reach, at least similar configurations might be found so that results can be extrapolated to solar wind conditions.

\section*{Acknowledgments}
\label{sec:acknowledgments}
The authors gratefully acknowledge the Gauss Centre for Supercomputing e.V. (www.gauss-centre.eu) for funding this project by providing computing time on the GCS Supercomputer SuperMUC at Leibniz Supercomputing Centre (LRZ, www.lrz.de).
This work is based upon research supported by the National Research Foundation and Department of Science and Technology.
Any opinion, findings and conclusions or recommendations expressed in this material are those of the authors and therefore the NRF and DST do not accept any liability in regard thereto.


\begin{thebibliography}{99}
\bibitem{kilian12} P. Kilian, T. Burkart, F. Spanier, \emph{The Influence of the Mass Ratio on Particle Acceleration by the Filamentation Instability} in \emph{High Performance Computing in Science and Engineering '11}, Springer, Berlin Heidelberg 2012.
\bibitem{lange13} S. Lange, F. Spanier, M. Battarbee, R. Vainio, T. Laitinen, \emph{Particle scattering in turbulent plasmas with amplified wave modes}, \emph{Astronomy and Astrophysics} {\bf553} (2013) A129 [{\tt1303.7463}].
\bibitem{schreiner14} C. Schreiner, F. Spanier, \emph{Wave-Particle-Interaction in Kinetic Plasmas}, \emph{Computer Physics Communications} {\bf185} (2014) [{\tt1404.0499}].
\bibitem{jokipii66} J. R. Jokipii, \emph{Cosmic-Ray Propagation. I. Charged Particles in a Random Magnetic Field}, \emph{Astrophysical Journal} {\bf146} (1966).
\bibitem{lee74} M. A. Lee, I. Lerche, \emph{Waves and Irregularities in the Solar Wind}, \emph{Reviews of Geophysics and Space Physics} {\bf12} (4) (1974).
\bibitem{schlickeiser89} R. Schlickeiser, \emph{Cosmic-ray transport and acceleration}, \emph{Astrophysical Journal} {\bf336} (1989).
\bibitem{michalek96} G. Micha{\l}ek, M. Ostrowsky, \emph{Cosmic ray momentum diffusion in the presence of nonlinear Alfv{\'e}n waves}, \emph{Nonlinear Processes in Geophysics} {\bf3} (1996).
\bibitem{qin01} G. Qin, W. H. Matthaeus, J. W. Bieber, \emph{Perpendicular Transport of Charged Particles in Composite Model Turbulence: Recovery of Diffusion}, \emph{Astrophysical Journal} {\bf578} (2001)
\bibitem{stix92} T. H. Stix, \emph{Waves in plasmas}, Springer, New York 1992.
\bibitem{hockney88} R. W. Hockney, J. W. Eastwood, \emph{Computer simulation using particles}, Hilger, Bristol 1988.
\bibitem{vainio03} R. Vainio, T. Laitinen, H. Fichtner, \emph{A simple analytical expression for the power spectrum of cascading Alfv\'en waves in the solar wind}, \emph{Astronomy and Astrophysics} {\bf407} (2003).
\end{thebibliography}
\end{document}